\documentclass{article}
\usepackage{comment}
\usepackage{color}
\usepackage{PRIMEarxiv}
\usepackage[utf8]{inputenc} 
\usepackage[T1]{fontenc}    
\usepackage{amsmath}

\usepackage{hyperref}       
\usepackage{url}            
\usepackage{booktabs}       
\usepackage{amsfonts}       
\usepackage{nicefrac}       
\usepackage{microtype}      
\usepackage{lipsum}
\usepackage{fancyhdr}       
\usepackage{graphicx}       
\graphicspath{{media/}}     

\title{Optimal Covariance Cleaning for Heavy-Tailed Distributions: Insights from Information Theory
}

\author{
  Christian Bongiorno\\
  Université Paris-Saclay, CentraleSupélec,\
 \\
 Laboratoire de Mathématiques et Informatique\\
 pour la Complexité et les Systèmes,\\
 91192 Gif-sur-Yvette, France.\\
  \texttt{christian.bongiorno@centralesupelec.fr} \\
   \And
  Marco Berritta \\
  Department of Physics and Astronomy, \\
  University of Exeter, Stocker Road,\\
  Exeter EX4 4QL, UK.
}

\begin{document}
\maketitle

\begin{abstract}

In optimal covariance cleaning theory, minimizing the Frobenius norm between the true population covariance matrix and a rotational invariant estimator is a key step. This estimator can be obtained asymptotically for large covariance matrices, without knowledge of the true covariance matrix. In this study, we demonstrate that this minimization problem is equivalent to minimizing the loss of information between the true population covariance and the rotational invariant estimator for normal multivariate variables. However, for Student's t distributions, the minimal Frobenius norm does not necessarily minimize the information loss in finite-sized matrices. Nevertheless, such deviations vanish in the asymptotic regime of large matrices, which might extend the applicability of random matrix theory results to Student's t distributions. These distributions are characterized by heavy tails and are frequently encountered in real-world applications such as finance, turbulence, or nuclear physics. Therefore, our work establishes a connection between statistical random matrix theory and estimation theory in physics, which is predominantly based on information theory.
\end{abstract}


\textit{Introduction.--} In today's data-rich environment, multivariate analysis has taken a central role across various fields, primarily due to the surge in computational capabilities and data availability. This shift necessitates the estimation of large covariance matrices, a process that is unfortunately susceptible to the "curse of dimensionality". This term describes the challenge that arises when the number of variables $n$ is large relative to the number of observations $t$. In such cases, the covariance matrix becomes increasingly noisy and can even turn non-positive definite when $n>t$. It's often the case that we must operate under conditions where $n$ is approximately equal to $t$ due to factors such as non-stationarity, placing intrinsic limits on our data collection capacity.

To address this issue, a frequently employed strategy is the correction of eigenvalues in the sample covariance matrix. The earliest of such methods, rooted in Random Matrix Theory (RMT), was proposed in Ref.\cite{laloux1999noise}. The fundamental proposition of this method is the convergence of the eigenvalue distribution of a random noise time series to the Marchenko-Pastur distribution when $n,t \to \infty$ with $q=n/t>1$ finite. Real-world matrices, like those representing daily returns in financial markets, partially conform to the Marchenko-Pastur distribution but also exhibit outlier eigenvalues. In the method from Ref.\cite{laloux1999noise}, all eigenvalues $\lambda$ that are less than a threshold $\lambda_{\max}$ are treated as sample noise fluctuations and are filtered out. Here, $\lambda_{\max}$ is the maximum eigenvalue predicted by the Marchenko-Pastur distribution. This value serves as a boundary between the bulk of the distribution, representing noise, and the outliers, potentially representing signal.

However, this approach  has been replaced by more effective techniques, proving that the sample eigenvalues that belong to the bulk of the Marchenko-Pastur carry relevant information. Starting from this observation, a large variety of approaches to address the optimal eigenvalue correction were proposed \cite{ledoit2011eigenvectors}. Needless to say, behind any idea of optimality, a proper covariance estimator target must be defined. The case of the former cited method is the oracle estimator \cite{bun2016rotational}. In a simple world, the oracle estimator is the optimal correction that minimizes the Frobenious norm distance of the unknown population matrix. As the population matrix is unknown, assumptions like large sample size limit and stationarity are required to have an asymptotic approximation for such estimators. The latter is, of course, one of the strongest, as shown in Ref.~\cite{bongiorno2023filtering}.

In this work, we propose a new approach to the target matrix. We aim to minimize the information lost by our estimator when approximating the population matrix. To quantify this information loss, we employ the Kullback-Leibler (KL) divergence \cite{joyce2011kullback}. For two probability measures $P$ and $Q$ over a set $\mathcal{X}$, the KL divergence is defined as:

\begin{equation}\label{eq:KLgen}
KL(P||Q) := \int_{\mathcal{X}} \left( \log \frac{P}{Q} \right) dP.
\end{equation}

This measure, a cornerstone of information theory, provides a quantification of the informational discrepancy between two probability distributions. It essentially measures the "informational cost" of approximating one distribution (in our case, the population matrix) with another (the estimator).

The utilization of the KL divergence is not confined to information theory. It has also found extensive application in various physical contexts. Indeed, the task of estimating physical quantities from noisy measurements is pervasive in physics, and methods that revolve around minimizing the KL divergence to extract information about a system have been widely employed. For example, it has been used to study spin-glasses \cite{leuzzi08}, in high-energy physics \cite{gambhir22}, to extract information from gravitational waves detection \cite{chua20}, and to estimate the electron density in solid-state systems \cite{de_vries96}. It has also been deployed in a host of other contexts \cite{dimario20,everett21}. Interestingly, the quantum extension of the KL divergence - the quantum relative entropy and its associated quantity, the quantum Fisher information \cite{vedral02}, has played a pivotal role in quantum metrology \cite{sidhu20}. This is not surprising, considering that one of the primary objectives of quantum metrology is to estimate a physical quantity based on a finite set of noisy measurements, a problem that our work also addresses. In general, the quantum relative entropy is central to several concepts in quantum mechanics. For instance, quantities such as entanglement or purity are often estimated using the trace distance or the fidelity, both of which are related to the KL divergence \cite{nielsen12}. In such cases, the loss of information is typically quantified using the Fisher information metric. However, the Fisher information metric's locality precludes its use in scenarios where very little is known beforehand \cite{braunstein94}. Furthermore, in quantum mechanics, covariance matrices are instrumental in characterizing the entanglement properties of multipartite systems \cite{adesso04}. Specifically, the covariance matrix can be leveraged to construct the logarithmic negativity, a measure of entanglement \cite{vidal02}.

The problem of estimating physical quantities from noisy measurements when the underlying probability distribution is a fat-tailed distribution, such as the Student's t-distribution, is common in various fields of physics~\cite{majda2014conceptual,biro2005power,barabasi2009scale,millhauser1988diffusion} and finance~\cite{thurner2012leverage,lillo2003power,sornette2000portfolio}. In such scenarios, a Gaussian assumption often fails to capture the real-world variability of the data.

In this context, the application of the Kullback-Leibler (KL) divergence for correlation matrix filtering for Normal multivariate variables was first introduced in Ref.~\cite{tumminello2007kullback}. A preliminary attempt to generalize this to multivariate heavy-tailed distributions was made in Ref.\cite{biroli2007student}, although it considered only a simple single factor model with heavy tails. To date, a closed expression of the KL divergence for multivariate Student's t-distributions remains elusive, with the work in Ref.\cite{contreras2014asymptotic} providing a closed expression under the quasi-normality assumption, and only for a t-student distributed probability.

In our work, we delve deeper into this problem. We explore the relationship between estimators based on the minimal Frobenius norm and the KL divergence, and how they behave differently for fat-tailed distributions. We highlight that, while the target estimators for the minimal Frobenius norm and KL coincide for normal multivariate variables, they diverge in the finite $n$ regime for fat-tailed distributions. Consequently, we develop a numerical approach for optimal correction for the Student's t-distribution and derive asymptotically the limiting equation for the KL divergence in the Student's t case in the large $n$ (thermodynamic) limit. In doing so, our work illuminates the interplay between the choice of the estimator and the underlying distribution, offering valuable insights into the estimation problem in situations where a Gaussian assumption is not suitable.

\textit{Problem Statement.--} The main goal of the filtering methods is to find the best Rotational Invariant Estimator (RIE) for the population correlation matrix ${\bf C}$. An RIE is 
\begin{equation}
    {\bf \Xi}({\bf \Lambda}) :={\bf V \Lambda V}',
\end{equation}
where ${\bf V} \in SO(n)$ are given, and they are not eigenvectors of $\bf{C}$ but, in general, are the eigenvectors of the sample correlation matrix. The standards correlation cleaning approaches rely on finding the eigenvalue matrix ${\bf\Lambda}_F$ that minimizes the Frobenious norm distance\footnote{The Frobenious norm distance is $\sum_{ij} \left(A_{ij}-B_{ij}\right)^2= \mbox{tr}\left[\left({\bf A} -{\bf B}\right)^2\right]$} with the population matrix $||{\bf \Xi}({\bf \Lambda}) - {\bf C}||_F$. Such estimator, when the population matrix ${\bf C}$ is known is called Oracle, and it can be obtained \cite{bun2016rotational} from  
\begin{equation}\label{eq:frob}
    {\bf \Lambda}_F = \left( {\bf V' C V}\right)_d,
\end{equation}
where the operator $(\bullet)_d$ sets to zero the out-of-diagonal elements.

The optimal RIE that minimizes the KL divergence assuming that the population matrix is known, must be derived from the multivariate form of the KL  
\begin{equation}\label{eq:KL}
    \mbox{KL}({\bf C}||{\bf \Xi}({\bf \Lambda})) :=    \mathcal{E}\left[ \log \left( \frac{ \mathcal{P}(x;{\bf C})}{ \mathcal{P}(x;{\bf \Xi}({\bf \Lambda}))}\right) \right]_{ \mathcal{P}(x;{\bf C})}.
\end{equation}

Here, the operator $\mathcal{E}$ stands for the expectation. In this formulation, the KL divergence represents the expected information loss when we use $\mathcal{P}(x;{\bf \Xi}({\bf \Lambda}))$ as an approximation for data that are actually distributed according to $\mathcal{P}(x;{\bf C})$. More specifically, it calculates the expected value of the logarithmic ratio of the probability density functions (PDFs) with covariance matrices ${\bf C}$ and ${\bf \Xi}({\bf \Lambda})$ for a multivariate random variable with zero means and population covariance matrix ${\bf C}$.

\textit{Multivariate Gaussian.–} For multivariate Gaussian variables with zero means, the KL divergence is already well-established and can be expressed as follows:
\begin{eqnarray}\label{eq:KLnorm}
   \mbox{KL}({\bf C}||{\bf \Xi}({\bf \Lambda})) = \frac{1}{2}\left[\mbox{Tr}\left[ {\bf \Xi}({\bf \Lambda})^{-1} {\bf C} \right] -n + \log \left( \frac{\mbox{det} \left[{\bf \Xi}({\bf \Lambda}) \right] }{\mbox{det} \left[{\bf C} \right] }\right)\right]
\end{eqnarray}
Interestingly, for Gaussian multivariate variables, the eigenvalues that minimize the Frobnious norm also minimize the Kullback-Leibler divergence, i.e. ${\bf \Lambda}_F={\bf \Lambda}_{KL}$.
The former result can be obtained by solving
\begin{equation}\label{eq:KLgauss}
    \partial_{\lambda_k} \mbox{KL}({\bf C}||{\bf \Xi}({\bf \Lambda})) = \frac{1}{2}\left(\frac{1}{\lambda_k} -\frac{1}{\lambda_k^2} \mbox{Tr}\left[ {\bf v}_k{\bf v}_k' {\bf C}  \right]\right) = \frac{1}{2}\left(\frac{1}{\lambda_k} -\frac{1}{\lambda_k^2}  {\bf v}_k' {\bf C} {\bf v}_k\right) =0.
\end{equation}

That leads to eq.~\eqref{eq:frob}. 

\textit{Multivariate t-student.--} If both distributions are instead two multivariate t-student random variables, the computation is more challenging and requires tailored approximations. The PDF for a standardized t-student of $n$ random variables is
\begin{equation}\label{eq:tdist}
   \mathcal{P}({\bf x} ; {\bf C},\nu) =   \frac {\Gamma \left((\nu +n)/2\right)}{\Gamma (\nu /2)\nu ^{n/2}\pi ^{n/2}\left|{ \bf{C}  }\right|^{1/2}} \left( 1+\frac{1}{\nu} \bf{x}'\bf{C}^{-1} \bf{x} \right)^{-\frac{n+\nu}{2}}.
\end{equation}

\begin{figure}[tbh]
    \centering
    \includegraphics[width=0.45\columnwidth]{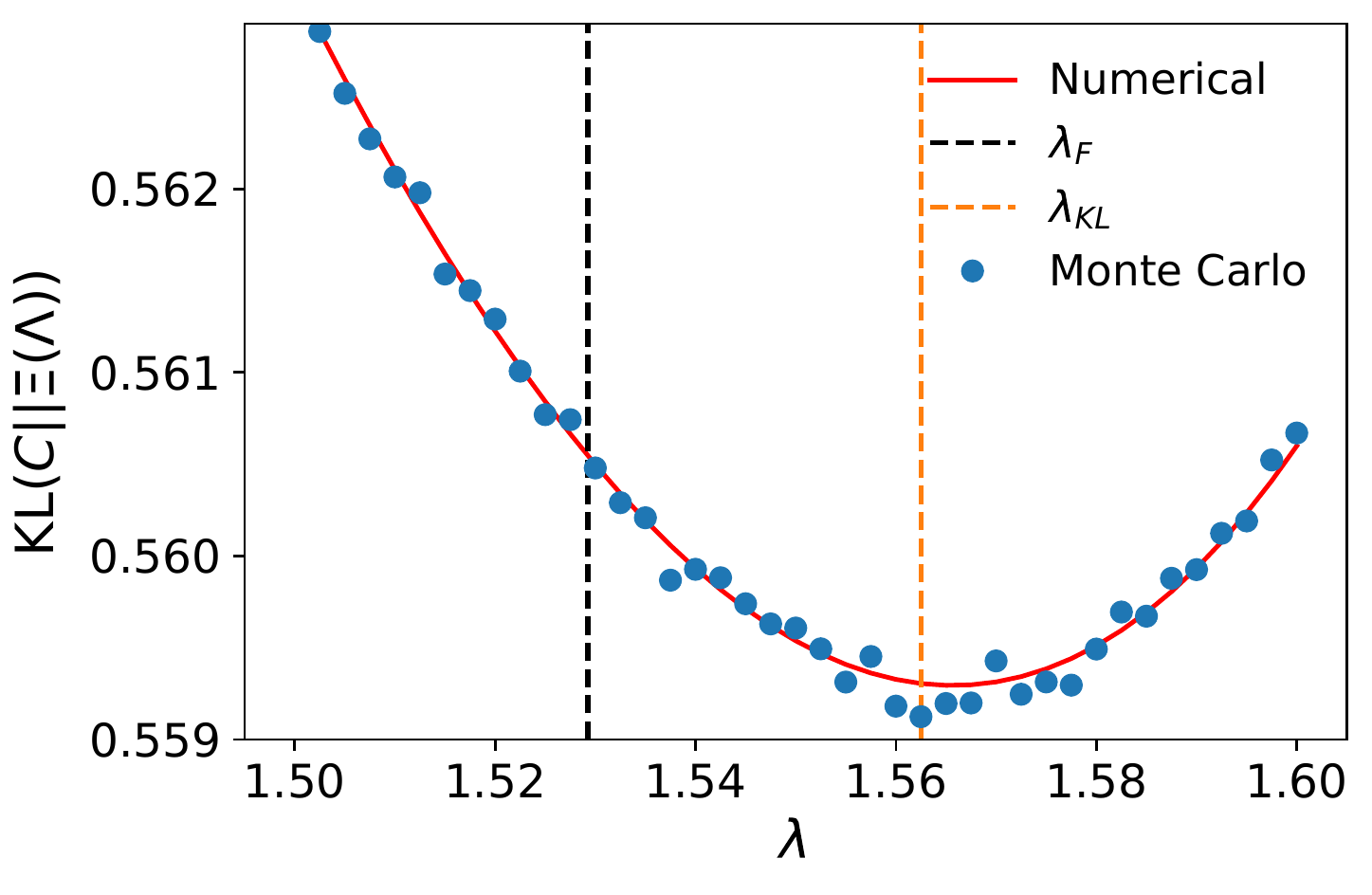}
    \includegraphics[width=0.45\columnwidth]{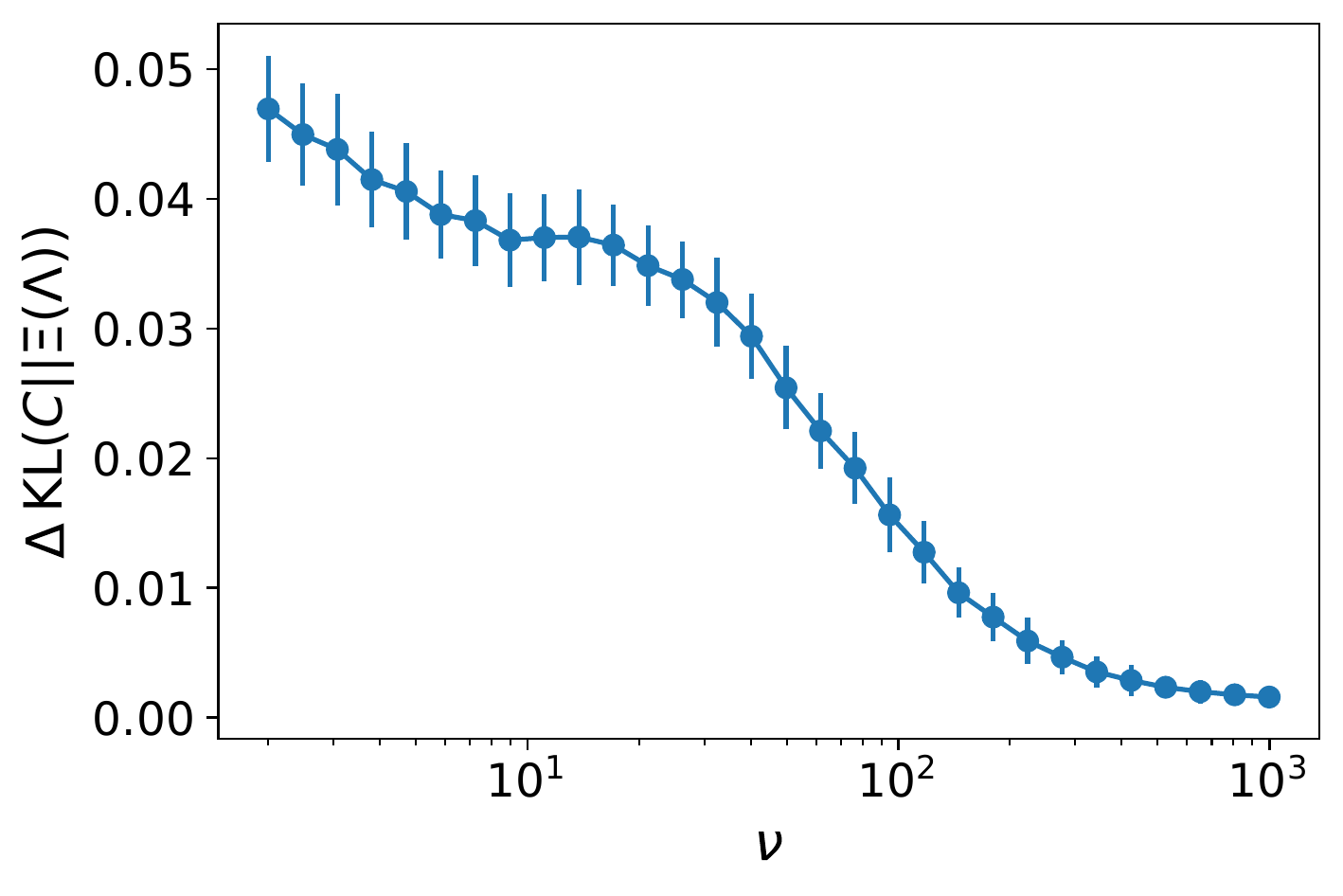}
    \caption{The left panel shows the KL for a matrix with $n=2$ and a student's t distribution with $\nu=4$, the plot reports the Monte Carlo estimation and the numerical quadrature obtained integrating the two variables in $[-100,100]$. The right panel shows the difference in KL between ${\bf \Xi}({\bf \Lambda}_F)$ and ${\bf \Xi}({\bf \Lambda}_{KL})$ for the student's t distribution with parameter $\nu$ of a synthetic covariance matrix ${\bf C}$ with $n=30$ and an eigenvalue distribution that comes from a geometric progression with exponent $1.8$. The eigenvector ${\bf V}$ is obtained by applying a small random rotation in terms of Euler angles to the original eigenvalues of ${\bf C}$. The optimal ${\bf \Lambda}_{KL}$ is obtained with a Montecarlo computed with $10,000$ samples. The plot indicates the mean and the standard deviation of $100$ runs.  }
    \label{fig:monte}
\end{figure}

Our first step is to illustrate the difference between the Gaussian and the student's t cases in a low sample size (n) scenario. For this, we have developed a Monte Carlo methodology, which can be found in the referenced repository \cite{code}, that calculates the expected value of multiple random realizations of eq.~\eqref{eq:KL} for random variables drawn from the distribution given in eq.~\eqref{eq:tdist}.
In the left panel of Fig.\ref{fig:monte}, we depict the difference between the minimum of the KL divergence and the Frobenius norm. In this particular example, we have used $n=2$, which means that the correlation's eigenvalue has just one degree of freedom, given that they must sum up to $n$. This figure also includes a numerical approximation of the KL divergence obtained through an integral quadrature method. However, this method becomes impractical as $n$ increases. For larger values of $n$, we can obtain a numerical approximation of $ \Delta \mbox{KL}$ by using a combination of Sequential Least Squares Programming (SLQP) minimization and the Monte Carlo approach. This numerical approximation accentuates the discrepancy with the Frobenius norm estimator as $\nu$ approaches 2, as demonstrated in Fig.\ref{fig:monte}

\textit{Asymptotic derivation.--} Given that a numerical discrepancy is observed for small $n$, our aim is to derive an asymptotic approximation of the KL for two Student's t-distributions in the large $n$ limit. This involves calculating the expected value of
\begin{equation}
\log \left( \frac{\mathcal{P}({\bf x} ; {\bf C},\nu) }{\mathcal{P}({\bf x} ; {\bf \Xi}({\bf \Lambda}),\nu) }\right) = \frac{1}{2}\left[\log \frac{|{\bf \Xi}({\bf \Lambda})|} {| {\bf C}|} + (n+\nu)   \log \left( \frac{1 + \frac{1}{\nu}\bf{x}'{\bf \Xi}({\bf \Lambda}^{-1})  \bf{x}}{1 + \frac{1}{\nu}\bf{x}'\bf{C}^{-1} \bf{x} } \right)\right]
\end{equation}`

with $x \sim  \mathcal{P}({\bf x} ; {\bf C},\nu)$. To compute the expected value of the first term, we leverage the linearity of the expectation operator and examine each logarithm of the ratio separately. The last term poses a greater challenge. We focus initially on the first-order two-variable Taylor expansion of the variance of the argument of the logarithm ratio. This is centered around the expected values of the quadratic bilinear forms ${{\bf x}'{\bf \Xi}(\bf \Lambda}^{-1}){\bf x}$ and ${\bf x}' {\bf C}^{-1} {\bf x}$, which we denote as $a$ and $b$ respectively,  and $c=1/\nu$. The detailed computations for the variance, the covariance, and the expectation in the multivariate t-distributions case are provided in Ref.~\cite{rong2012quadratic}.  This approach leads us to

\begin{eqnarray}
\mathcal{V}\left[ \frac{ 1 + c a}{1 +c b} \right] \approx c^{2}\left[\frac{\mathcal{V}\left[a\right]}{\left(1+c \mathcal{E}\left[b\right]\right)^2}-2 \frac{1+c \mathcal{E}\left[a\right]}{\left(1+c \mathcal{E}\left[b\right]\right)^3}\mathcal{C}\left[a,b\right] +\frac{\left(1+c \mathcal{E}\left[a\right]\right)^2}{\left(1+c \mathcal{E}\left[b\right]\right)^4}\mathcal{V}\left[b\right] \right] = \nonumber \\ =
 \frac{2 (\nu -2) n \left(\nu -n \overline{\mbox{Tr}[\bf{C  \Xi}({\bf \Lambda}^{-1})] }^2+\overline{\mbox{Tr}[\bf{C  \Xi}({\bf \Lambda}^{-1})\bf{C  \Xi}({\bf \Lambda}^{-1})] } (\nu +n-2)-2 (\nu -2) \overline{\mbox{Tr}[\bf{C  \Xi}({\bf \Lambda}^{-1})] }-2\right)}{(\nu -4) (\nu +n-2)^3}
\end{eqnarray}

where we extracted the system size dependence from the traces $\mbox{Tr}[{\bf C  \Xi}({\bf \Lambda}^{-1})] = n  \overline{\mbox{Tr}[\bf{C  \Xi}({\bf \Lambda}^{-1})] }$,  and $\mbox{Tr}[\bf{C  \Xi}({\bf \Lambda}^{-1})\bf{C  \Xi}({\bf \Lambda}^{-1})] = n \overline{\mbox{Tr}[\bf{C  \Xi}({\bf \Lambda}^{-1}){\bf C  \Xi}({\bf \Lambda}^{-1})] }$. In the large system limit

\begin{equation}
 \lim_{n \to \infty} \mathcal{V}\left[ \frac{ 1+\frac{1}{\nu} {\bf x}'{\bf \Xi}({\bf \Lambda}^{-1}){\bf x}}{ 1+\frac{1}{\nu}{\bf x}' {\bf C}^{-1} {\bf x}} \right] = 0  
\end{equation}
As a result, in the large n limit, the distribution of the arguments becomes a delta distribution centered in their expected values, which simplifies and allows to compute the large n approximation of the KL as

\begin{equation}
    \mbox{KL}({\bf C}||{\bf \Xi}({\bf \Lambda})) \approx \frac{1}{2}\left\{ \log \frac{ |{\bf \Xi}({\bf \Lambda})| }{ | {\bf C}|}+ (n+\nu)\left[\log \left(1 +  \frac{n \overline{\mbox{Tr}[\bf{C  \Xi}({\bf \Lambda}^{-1})] }}{\nu-2} \right) - \log \left(1 +  \frac{n}{\nu-2} \right)\right]\right\}
\end{equation}
Then the normalized KL can be derived with an asymptotic limit
\begin{equation}\label{eq:KLtstud}
\overline{\mbox{KL}({\bf C}||{\bf \Xi}({\bf \Lambda}))} = \lim_{n \to \infty}\frac{\mbox{KL}({\bf C}||{\bf \Xi}({\bf \Lambda}))}{n} = \frac{1}{2} \left(  \overline{ \log |{\bf \Xi}({\bf \Lambda})|}-\overline{\log | {\bf C}|}  + \log \overline{\mbox{Tr}[\bf{C  \Xi}({\bf \Lambda}^{-1})] } \right)
\end{equation}
and it is independent of $\nu$.  From the former equation, it is possible to obtain the Gaussian case with the limit 
\begin{equation}
     \lim_{\nu \to \infty}\frac{\mbox{KL}({\bf C}||{\bf \Xi}({\bf \Lambda}))}{n} = \frac{1}{2} \left( \overline{ \log |{\bf \Xi}({\bf \Lambda})|}-\overline{\log | {\bf C}|} +  \overline{\mbox{Tr}[\bf{C  \Xi}({\bf \Lambda}^{-1})] } -1 \right).
\end{equation}
The derivative of the normalized KL of eq.\eqref{eq:KLtstud} in the eigenvalues by expressing n again, that in the large limit is approximately
\begin{equation}
   \partial_{\lambda_k} \overline{\mbox{KL}({\bf C}||{\bf \Xi}({\bf \Lambda}))}\approx \frac{1}{2}\left(\frac{1}{n \lambda_k} - \frac{1}{\lambda_k^2} \frac{ {\bf v}_k' {\bf C} {\bf v}_k}{\mbox{Tr}[{\bf C  \Xi}({\bf \Lambda}^{-1})]} \right),
\end{equation}
which is equal to zero in the $\Lambda_F=({\bf V}'{\bf C V})_d$ since $\mbox{Tr}[{\bf C  \Xi}({\bf \Lambda}_F^{-1})]=n$. As a result, the former equation has the same zeros of  eq.~\eqref{eq:KLgauss}. Proving the equivalence of the minimum for the Gaussian and students t cases in the large system limit.

Another interesting observation is that in the $n$ large limit also the normalized KL of eq.~\eqref{eq:KLtstud} in the $\Lambda_F$ is equal for the Gaussian and student's t cases to
\begin{equation}\label{eq:EqOracle}
 \overline{\mbox{KL}({\bf C}||{\bf \Xi}({\bf \Lambda_F}))}=  \frac{1}{2} \left( \overline{ \log |{\bf \Xi}({\bf \Lambda_F})|}-\overline{\log | {\bf C}|} \right) 
\end{equation}
Finally, if $\nu$ is not a negligible fraction of $n$ we could write $\nu= h n$ with $h$ finite and non-zero.  Then
\begin{equation}\label{eq:KLh}
\lim_{n \to \infty}\frac{\mbox{KL}({\bf C}||{\bf \Xi}({\bf \Lambda}))}{n} = \frac{1}{2} \left[  \overline{ \log |{\bf \Xi}({\bf \Lambda})|}-\overline{\log | {\bf C}|}  + (1+h)\log \left( h + \overline{\mbox{Tr}[\bf{C  \Xi}({\bf \Lambda}^{-1})] } \right) -(1+h) \log (1+h) \right]
\end{equation}

The former equation highlights that a discrepancy between the Normal and Student' t case is observed for small h, while the discrepancy disappears for $h \to \infty$ or if the equation is computed in $\Lambda_F$.

\begin{figure}[tbh]
    \centering
    \includegraphics[width=0.45\columnwidth]{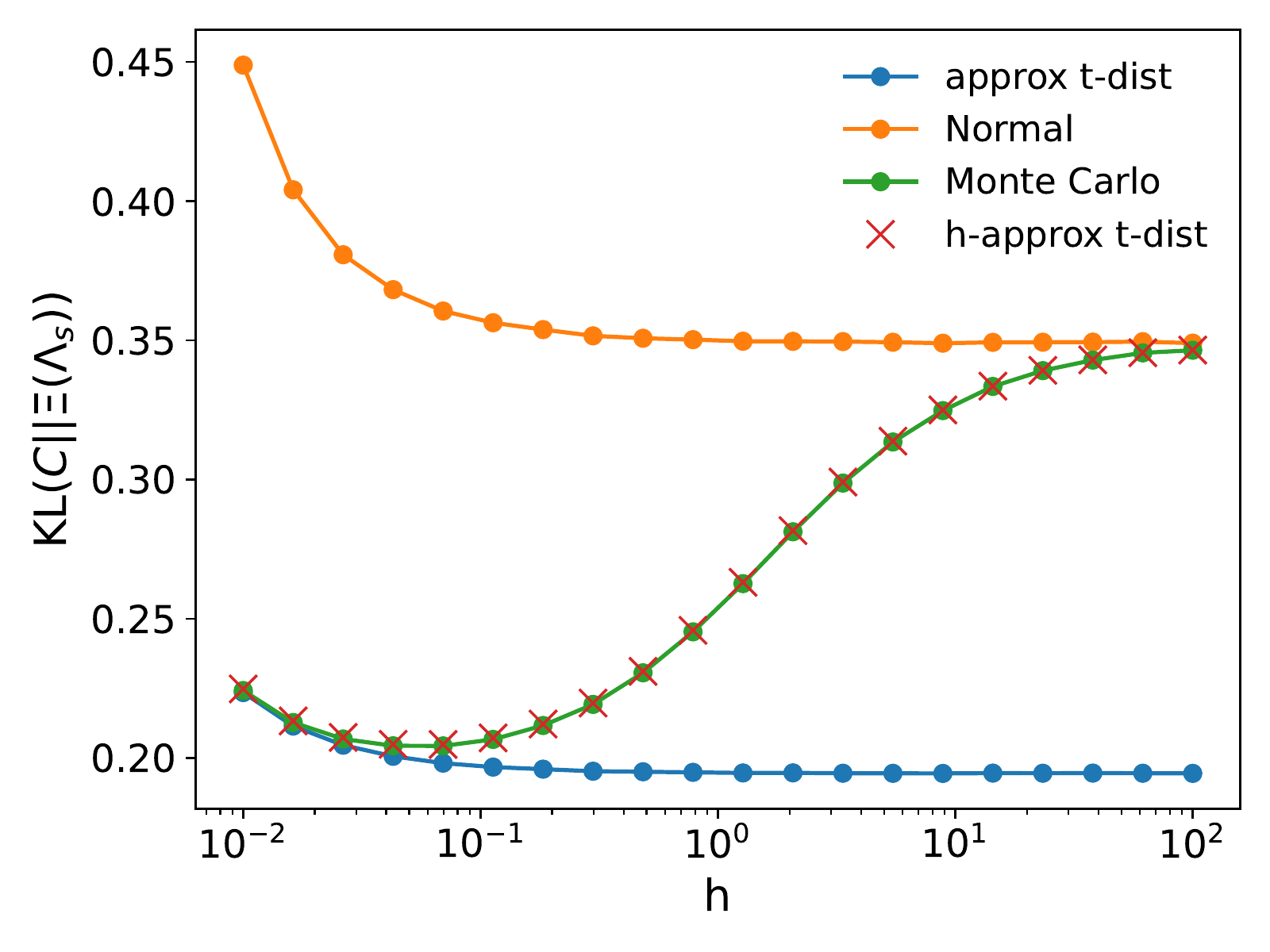}
    \includegraphics[width=0.45\columnwidth]{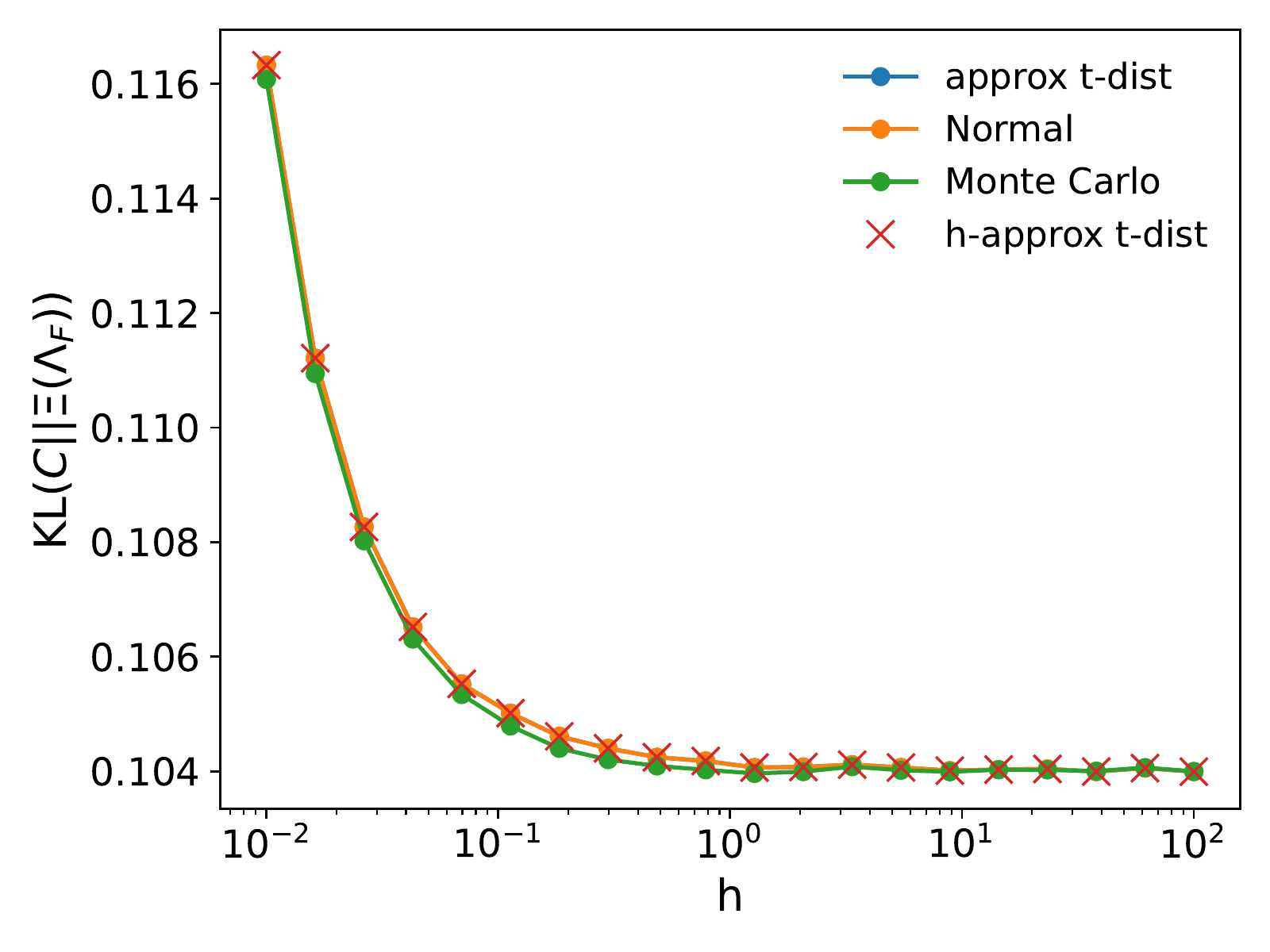}
    \caption{The left panel shows simulations with $n=1000$ of the KL between the population and the sample covariance matrix as a function of $h$; the right plot shows the KL between the population matrix and the Oracle estimator. The different lines represent, eq.~\eqref{eq:KLtstud} (blue), eq.~\eqref{eq:KLnorm} (orange), the numerical Monte Carlo (green) and eq.~\eqref{eq:KLh} . The values are averaged over 100 independents run.}
    \label{fig:KLasympt}
\end{figure}

In Fig.\ref{fig:KLasympt} we show that our approximation converges pretty well to the asymptotic expectations for moderately large $n=1000$. In particular, on the left plot, we confirm that the deviation from the Normal expectations is observed also for very large values of $\nu$, whenever the ratio $h=\nu/n$ is not large. In the right plot, we confirm eq.~\eqref{eq:EqOracle}, in fact, all estimates converge to the same values independently from the value of $h$.  

\textit{Discussion.--} In summary, we have demonstrated that the key step of optimal covariance cleaning theory, namely minimizing the Frobenius norm between the true population covariance matrix and a rotational invariant estimator, is equivalent to minimizing the loss of information between the true population covariance and the rotational invariant estimator for normal multivariate variables. We have shown that this equivalence does not necessarily hold for Student's t distributions in finite-sized matrices, but that it holds asymptotically for large matrices. This result might help to extend the applicability of random matrix theory to Student's t distributions, which are commonly encountered in real-world applications such as finance.

Furthermore, our work establishes a connection between statistical random matrix theory and estimation theory in physics, which is primarily based on information theory. The use of information theory has been instrumental in the development of a wide range of physical theories and models, such as the maximum entropy principle, which has been used to derive equilibrium thermodynamics from information theory. Our findings suggest that information theory can also provide valuable insights in the field of optimal covariance cleaning theory, which has important applications in statistical data analysis, signal processing, and machine learning.

In future work, it would be interesting to explore the applicability of our results to other heavy-tailed distributions and to investigate whether the use of alternative metrics for quantifying the loss of information, such as the Kullback-Leibler divergence, could lead to improved performance in finite-sized matrices. Additionally, it would be valuable to study the performance of optimal covariance cleaning in the presence of missing or incomplete data, which is a common issue in many real-world applications. Overall, our work highlights the potential of combining concepts from information theory and random matrix theory to develop more robust and accurate statistical methods for analyzing complex data sets.

\section*{Acknowledgments}
We thank Damien Challet and Lamia Lamrani for their helpful discussions.

\bibliography{references}
\bibliographystyle{unsrt}

\end{document}